# Terahertz emission from anomalous Hall effect in a single-layer ferromagnet

Qi Zhang, [1,2] Ziyan Luo,[1] Hong Li, [2,3] Yumeng Yang,[1] Xinhai Zhang,[2,*] and Yihong Wu[1,†]

[1] *Department of Electrical and Computer Engineering, National University of Singapore, 4 Engineering Drive 3, Singapore 117583, Singapore*

[2] *Department of Electrical & Electronic Engineering, Southern University of Science and Technology, Xueyuan Rd 1088, Shenzhen 518055, China*

[3] *Institute of Applied Physics and Material Engineering, University of Macau, Avenida da Universidade, Macau 999078, China*

We report on terahertz emission from a single layer ferromagnet which involves the generation of backflow nonthermal charge current from the ferromagnet/dielectric interface by femtosecond laser excitation and subsequent conversion of the charge current to a transverse transient charge current via the anomalous Hall effect, thereby generating the THz radiation. The THz emission can be either enhanced or suppressed, or even the polarity can be reversed, by introducing a magnetization gradient in the thickness direction of the ferromagnet. Unlike spintronic THz emitters reported previously, it does not require additional non-magnetic layer or Rashba interface.

Corresponding authors: Email: *zhangxh@sustc.edu.cn, †elewuyh@nus.edu.sg

# I. INTRODUCTION

Recently spin-to-charge conversion in femtosecond (fs) laser excited magnetic heterostructures has attracted attention as a promising mechanism for producing terahertz (THz) wave with magnetically controllable polarization state [1]. The key to the THz emission is the generation of spin-polarized super-diffusive charge current from a ferromagnetic layer by fs laser excitation and subsequent conversion of the spin polarized current to a transverse charge current by either inverse spin Hall effect (ISHE) [1-10] or inverse Rashba-Edelstein effect (IREE) [11-13]. The former involves a ferromagnet (FM) / non-magnet (NM) heterostructure wherein when an fs laser is irradiated on the FM/NM heterostructure, non-equilibrium electrons are excited to the states above Fermi level, generating a super-diffusive spin polarized current normal to the interface [14,15]. When the spin polarized current enters the NM layer with large spin-orbit coupling (SOC), it is converted to a charge current in the transverse direction by ISHE, thereby generating the THz wave [1,2]. On the other hand, the THz emitter based on IREE typically consists of a FM with an adjacent Rashba interface, *e.g.*, FM/Ag/Bi [16]; in this case, the super-diffusive spin polarized current launched by the FM layer is converted to transverse charge current at the Ag/Bi interface via the IREE [11-13], which in the same way as the ISHE, generates the THz emission. In this letter, we examine the possibility of using anomalous Hall effect (AHE) [17] to generate THz emission in a single layer FM with a large SOC and demonstrate efficient THz emission in samples with a magnetization gradient.

As shown schematically in Fig. 1, when a thin metallic film is irradiated by an fs laser, the electrons are excited to states above the Fermi level (Fig. 1(a)). Immediately after the excitation, equilibration takes place through two dominant mechanisms, *i.e.*, electron-electron and electron-phonon interactions [18]. Due to the much smaller heat capacity, the electron subsystem quickly reaches a high temperature ($T_e$) Fermi-Dirac distribution within 0.1 ps (Fig. 1(b)), whereas the lattice stays close to the ambient temperature ($T_p$) [19]. The electrons subsequently cool and thermalize with its own lattice within a few picoseconds (Fig. 1(c)). In addition to phonons, the magnon also plays a role in the equilibration process of hot electrons in ferromagnets. The magnon temperature ($T_m$) is typically higher than that of phonons before all the three subsystems reach the thermal equilibrium [20,21].

Before the electron subsystem reaches the equilibrium (*i.e.*, $t < 0.1$ ps), the nonthermal electrons move at a fast speed (~$10^6$ m/s) in a super-diffusive manner (Fig. 1(d)) [14,15]. For electrons at a distance of at least a mean-free-path ($\lambda_e$) away from the top (*e.g.*, with MgO capping layer) and bottom (*e.g.*, with quartz substrate) interfaces, they will collide with other electrons to reach the thermal equilibrium within the electron subsystem. However, for electrons with a distance shorter than $\lambda_e$ from the interface, they will be reflected back to form a backflow current due to reflection at the FM/dielectric interface [15]. The amplitude of the backflow current depends strongly on the material properties and roughness of the interface. For metallic films deposited on a smooth substrate, typically the bottom interface will be smoother than the top interface. In this case, the backflow currents from the bottom ($j_{I1}$) and ($j_{I2}$) interfaces will not be completely canceled out, leading to a net backflow current ($j_{I1} - j_{I2}$). We argue that the backflow current will be converted to a transverse transient current when the FM has a large AHE. Recently, we have demonstrated that $(Fe_xMn_{1-x})_yPt_{1-y}$ thin films with optimum compositions exhibit a sizable AHE, and the backflow of spin accumulation induced by the AHE gives the anomalous Hall magnetoresistance (AHMR) [22]. Here, we show that $(Fe_xMn_{1-x})_yPt_{1-y}$ is also a promising material for AHE-based THz emitter. We substantiate our arguments by examine both the thickness and pumping fluence-dependence of the emission efficiency in $(Fe_xMn_{1-x})_yPt_{1-y}$ emitters. We show that the use of Pt composition gradient can either enhance or suppress the THz emission, depending on the gradient direction relative to normal direction of the two interfaces.

## II. EXPERIMENTAL RESULTS AND DISCUSSION

### A. AHE origin of THz emission

The sample preparation and THz measurement methods are given in the Supplemental Material [23,24]. Fig. 2(a) illustrates the THz generation from fs laser pumped single layer FM with either the quartz (up) or MgO (down) side pumping. We first measured the THz emission from a FeMnPt (3) thin film capped by MgO (4) (hereafter the number inside the brackets indicate film thickness in nm), and the results are shown in Fig. 2(b), where the upper and lower pair of waveforms are obtained from quartz side and MgO side pumping at a fluence of 555 μJ/cm$^2$, respectively (hereafter, FeMnPt refers to (Fe$_{0.8}$Mn$_{0.2}$)$_{0.67}$Pt$_{0.33}$). The solid and dashed lines correspond to the applied field in +$y$ and –$y$ direction, respectively. It is apparent that the THz polarity is reversed when either the applied magnetic field or pumping direction is reversed; the former indicates that the THz emission is of magnetic origin [25-27] whereas the latter implies that there is a reversal of the sign of the driving force that is responsible for the generation of the THz wave. Similar sign dependence on pumping direction has been observed in FM/NM bilayer emitters [2,6,8]. But, in those cases, it is understood that the sign reversal of THz wave polarization is caused by the reversal of the direction of spin polarized current due to sample flipping [3,10]. In the present case, however, there is only a thin layer of FM; the flipping of the sample will not break the symmetry as long as the sample is homogenous in the thickness direction, which is presumably the present case. Therefore, the most likely cause for the sign reversal is the presence of a net longitudinal current in the film thickness direction as discussed above. With presence of AHE, the longitudinal current is converted to a transient transverse current $j_t = \theta_{AHE} m \times j_l$ with $m$ the magnetization direction, $j_l$ the net longitudinal current, and $\theta_{AHE}$ the AHE angle. Since $j_l$ is physically related to two interfaces, the reverse of pumping direction, *i.e.*, sample flipping, will lead to the sign reversal of $j_l$. This explains why the polarity of THz reverses when either the pumping direction or magnetization direction is reversed. In addition to FeMnPt, we also conducted similar experiments by replacing FeMnPt with $Fe_{0.8}Mn_{0.2}$ (3), $Co_{0.2}Fe_{0.6}B_{0.2}$ (3), $Ni_{0.8}Fe_{0.2}$ (3) and the rest remains the same. Indeed, we observed THz radiations from all the three samples, though their amplitudes are much smaller compared to that of FeMnPt. Interestingly, the polarity of THz wave also reverses when either the pumping direction or the magnetic field direction is reversed for all the samples. We plot the peak values of THz waveforms of the four emitters in Fig. 2(c) with a positive magnetic field. The positive (negative) values correspond to quartz (MgO) side pumping. The NiFe shows a larger amplitude compared to FeMn and CoFeB, which is in agreement with the observation of a larger AHE in NiFe [28-30], though the roughness of the interface may also play a role. The above results provide strong evidences that the THz emission from FeMnPt is originated from the AHE.

### B. Thickness dependence of THz emission

To further test the relevance of the proposed AHE origin, we studied the thickness dependence of THz emission from single layer FeMnPt from 3 nm to 9 nm. Figs. 3(a) and 3(b) show the emission waveforms (shifted along the time axis for clarity) and the corresponding peak amplitude (symbol), respectively. As can be seen, the THz amplitude increases from 3 nm to 4 nm and then decreases monotonically as the thickness increases further, with a broad maximum at around 4 nm. The thickness dependence can be modeled by involving the following sub-processes: i) generation of non-equilibrium electrons by laser excitation, ii) generation of backflow current at the two interfaces, iii) creation of transverse transient charge current via AHE, and iv) conversion of the charge current to THz emission. Without losing generality, we assume that the electron reflection coefficients at the quartz/FM and FM/MgO interfaces are $r_1$ and $r_2$, respectively (Fig. 1(d)). We also assume that the resulted super-diffusive current decays exponentially from the two interfaces. Based on this model, the electric field of the THz emission may be written as:

$$E_{THz} \propto M_s(d) \cdot FA(d) v_e(d) \cdot \int_0^d \theta_{AHE} (r_1 e^{\frac{-x}{\lambda_e/2}} e^{-\frac{d-x}{\lambda_T}} - r_2 e^{\frac{x-d}{\lambda_e/2}} e^{-\frac{d-x}{\lambda_T}}) dx \cdot \frac{1}{n_{air} + n_{sub} + Z_0 \sigma d}$$

$$= M_s(d) \cdot FA(d) v_e(d) \cdot \theta_{AHE} [r_1 \lambda_1 e^{-\frac{d}{\lambda_T}}(1 - e^{-\frac{d}{\lambda_1}}) - r_2 \lambda_2 (1 - e^{-\frac{d}{\lambda_2}})] \cdot \frac{1}{n_{air} + n_{sub} + Z_0 \sigma d} \quad (1)$$

where $d$ is the FM thickness, $F$ the fluence, $A(d)$ the absorptance of the FM layer, $r_1$, $r_2$ the reflection coefficient at two interfaces, $\lambda_e$ the non-equilibrium electron mean-free-path, $\lambda_T$ the decay length of THz emission inside the FM layer, $v_e(d)$ the average electron speed, $M_s(d)$ the saturation magnetization, $n_{air}$ ($n_{sub}$) the reflective index of air (substrate), $Z_0$ the impedance of free space, and $\lambda_1 = \frac{\lambda_e \lambda_T / 2}{\lambda_T - \lambda_e / 2}$, $\lambda_2 = \frac{\lambda_e \lambda_T / 2}{\lambda_T + \lambda_e / 2}$. Here, we have included the explicit thickness dependence of $A$, $v_e$ and $M_s$ in the equation. $A(d)$ is obtained from fitting to the experimental data, which yields $A(d) = 0.0314d + 0.1767$ (with $d$ in nm). $v_e(d)$ is estimated from the electron temperature ($T_e$), which itself is related to the absorbed laser energy per unit volume ($E_a$) [31,32]: $E_a = \frac{\xi(T_e^2 - T_0^2)}{2}$, where $\xi$ is the electronic specific heat and $T_0$ the initial electron temperature (i.e., room temperature). Substitute $E_a = A(d)F/d$ into previous equation, one obtains $v_e \propto \sqrt{T_e} = (\frac{2A(d) \cdot F}{\xi d} + T_0^2)^{1/4}$, where $F$ is the laser fluence. In addition to $A$ and $v_e$, the saturation magnetization also depends on $d$, which drops quickly with reducing $d$ when $d < 5$ nm, as revealed in our previous study [22]. A polynomial fitting to the experimental data is used to calculate $E_{THz}$ in Eq.(1) (see Supplemental Material [23]). The first and second term in the integrand of Eq. (1) represents the longitudinal to transverse current conversion efficiency as well as small absorption of the THz wave by the metal layer near the two interfaces. The last term of Eq. (1) denotes the transverse current to THz conversion efficiency. As shown by the solid-line in Fig. 3(b), the thickness dependence of THz peak amplitude can be fitted well using Eq. (1) with the following parameters: $\lambda_e$ = 1.2 nm, $\lambda_T$ = 14 nm, $r_1$ = 0.6, $r_2$ = 0.1 (see Supplemental Material for more details [23]), $F$ = 555 μJ/cm$^2$ (experimental value), $\xi$ = 0.7 mJ/cm$^3$K$^2$, $n_{air}$ = 1, $n_{sub}$ = 1.453, $Z_0$ = 377 Ω and $\sigma$ = 5.32×10$^6$ S·m$^{-1}$. As the electron specific heat of FeMnPt is not available, we use the reported value for Fe [32] instead. Assuming every single photon excites an electron-hole pair, at an absorptance of 0.4, a laser beam with a fluence of $F$ = 555 μJ/cm$^2$ will lead to a nonthermal electron density of about 0.88 nm$^{-3}$, which is equivalent to an average electron spacing of 1.05 nm. Therefore, the fitted value of $\lambda_e$ = 1.2 nm is reasonable. The fitted THz decay length of 14 nm is comparable to the reported value in FM/NM bilayers [8]. The conductivity of 5.32 × 10$^6$ S·m$^{-1}$ is obtained experimentally for the 9 nm FeMnPt [22]. The good agreement between fitted and measured thickness dependence provides further support to the proposed AHE origin of THz emission.

### C. Pump fluence dependence of THz emission

Next, we discuss the pumping fluence dependence of THz radiation from the FeMnPt ($d$) emitter as shown in Fig. 4. The symbols denote the measured values and solid lines are fitting curves according to Eq. (1). At a fluence of 555 μJ/cm$^2$, $\lambda_1 \approx \lambda_1 \approx \frac{\lambda_e}{2}$ = 0.6 nm. Therefore, the fluence dependence of $E_{THz}$ is mainly determined by $M_s F v_e \lambda_e$ and the last term of Eq. (1) with $v_e \propto (A(d)/\xi d)^{1/4}(F+\Gamma)^{1/4}$ where $\Gamma = \frac{\xi d}{2A(d)}$ (note: $T_e \gg T_0$) and $\lambda_e \propto [A(d)F]^{-1/3}$. The laser heating induced magnetization reduction may be approximated by $M_s(d) = M_{s0}(d)[1-\gamma(T_e-T_0)]$, where $M_{s0}(d)$ is the saturation magnetization at $T_0$ and $\gamma$ a thickness-dependent constant. It is thickness-dependent because the temperature-dependence of magnetization is different at different thickness when $d$ is small. As shown in Fig. 4, the experimental results can be fitted very well using Eq. (1) by treating $\gamma$ as a thickness dependent fitting parameter. The inset shows the thickness dependence of $\gamma$ used in the fitting. As expected, $\gamma$ decreases with increasing $d$ as the heating effect becomes weaker at larger thickness (see more discussion in the Supplemental Material [23]) [33-35].

## III. DISCUSSION ON ENHANCEMENT OF TERAHERTZ EMISSION

With the confirmation of AHE origin, now we discuss how we can enhance the THz emission. An obvious approach is to optimize the Pt composition so as to increase the difference between AHE near the two interfaces, thereby increasing the THz efficiency. As discussed in the Supplemental Material [23], at a thickness of 3 nm, the THz peak-to-peak amplitude ratio of FeMnPt emitters deposited at a Pt power of 15, 35 and 50 W is 14.1:1.7:1, with the sample deposited at 15 W showing the strongest THz emission. This prompted us to introduce a Pt gradient in the thickness direction. When the film is irradiated by a laser, immediately after the laser excitation, the spin polarization of the non-equilibrium electrons is assumed to be the same as those in the equilibrium state at Fermi level [14]. Therefore, the Pt composition gradient will be converted to a spin chemical potential gradient $\nabla \boldsymbol{\mu}_s$ in the thickness direction [20,36]. With the presence of $\nabla \boldsymbol{\mu}_s$, the total transverse current may be written as $\boldsymbol{j}_t = \theta_{AHE} \boldsymbol{m} \times \boldsymbol{j}_{bf} + \theta_{SHE} \sigma \boldsymbol{s} \times \nabla(\mu_s/2e)$, where $\theta_{SHE}$ is the spin Hall angle, $\boldsymbol{s}$ the net spin polarization direction of non-equilibrium electrons and $\boldsymbol{\mu}_s$ the spin chemical potential [37,38]. Therefore, the introduction of Pt composition gradient may either enhance or reduce the THz emission, depending on the sign of $\nabla \boldsymbol{\mu}_s$ and relative contributions from the two terms.

Fig. 5(a) shows the THz waveform from FeMnPt (9) with uniform Pt composition (15 W),

which is almost the same as that of FeMnPt (3) shown in Fig. 2(b). Fig. 5(b) shows THz emission from FeMnPt (9) with positive Pt gradient (*i.e.*, Pt power increases from 15 W to 50 W during deposition). Compared to the waveform shown in Fig. 5(a), the polarity remains the same but the peak-to-peak amplitude is increased by a factor of 5.4 (212 versus 39 in arbitrary unit). This means that a positive Pt composition gradient helps to enhance the THz emission. On the other hand, as shown in Fig. 5(c), the sample with a negative Pt composition gradient (*i.e.*, Pt power decreases from 50 W to 15 W during deposition) gives a much smaller THz signal with a peak-to-peak amplitude of – 84. The negative sign indicates that the polarity is reversed compared to the previous two cases. The ratio of the peak-to-peak amplitudes of the three samples is 1:5.4:-2.15. If we simply estimate the peak-to-peak amplitude ratio of the three samples by using $r_1 = 0.6$, $r_2 = 0.1$ and the amplitude ratio from samples with uniform Pt composition as discussed above, the ratio of these samples turned out to be 1:1.18:-0.11. Obviously the two sets of values don't tally, suggesting that the magnetization gradient plays an important role. The discrepancy can be resolved by taking into account the contribution from $\nabla \boldsymbol{\mu}_s$, as discussed in Supplemental Material [23]. These results demonstrate that the spin chemical potential gradient plays an important role in THz emission from the single layer FM. With further optimization of the materials and composition gradient, the current approach can potentially lead to more efficient THz emission.

## IV. CONCLUSIONS

In summary, we have demonstrated an alternative mechanism for generating THz emission from ultrathin FM layer via the AHE, though the presence of AHE in the THz regime has been reported before [39-41]. The process involves generation of backflow super-diffusive current at the FM/dielectric interfaces and subsequent conversion of the charge current to transverse current via AHE, and thereby generating the THz radiation. The THz generation is mainly caused by the non-thermal super-diffusive current near the two interfaces; the contribution from laser excitation induced hot carrier gradient inside the sample, if any, should be very small considering the large effective penetration depth (see Supplemental Material [23]) compared to the film thickness [42]. Further investigations can be carried out in future using multiple wavelength excitations when the laser sources are available [43,44]. There is tremendous room for improvement of emission efficiency once materials with large AHE are found. In the meantime, as we demonstrated in this paper, the emission efficiency can also be enhanced by introducing a gradient in the FM magnetization.

## ACKNOWLEDGEMENTS

Y.H.W. would like to acknowledge support by Ministry of Education, Singapore under its Tier 2 Grants (grant no. MOE2017-T2-2-011 and MOE2018-T2-1-076). X.H.Z. would like to acknowledge the project by Shenzhen Peacock Plan (grant no. KQTD2015071710313656) and the project of Shenzhen Science and Technology Innovation Committee (grant no. JCYJ20160301114759922).

spintronic trilayer, Appl. Phys. Lett. **114**, 041107 (2019).

# FIGURE CAPTIONS

FIG. 1. (a)-(c) Sub-processes at different time scale after ultrafast laser excitation: (a) Excitation of non-equilibrium electrons. (b) High-temperature Fermi-Dirac distribution of electron subsystem. (c) Nearly thermal equilibrium state among electron, phonon and magnon. (d) Schematic of nonthermal electron reflection at quartz/FM and FM/MgO interfaces and the formation of backflow current.

FIG. 2. (a) Schematics of THz emission setup with pumping from the quartz (up) and MgO (down) side, respectively. (b) THz wave from FeMnPt. The upper pair of waveforms is from quartz side pumping and the lower pair is from MgO side pumping. Solid and dashed-lines correspond to positive and negative fields, respectively. (c) THz peak values from different emitters: FeMnPt, FeMn, CoFeB and NiFe, with positive applied H field (positive amplitude: quartz side pumping; negative amplitude: MgO side pumping).

FIG. 3. Thickness dependence of THz waveform from FeMnPt single-layer FM emitter: (a) THz waveform at different FeMnPt thickness (shifted in time axis for clarity) and (b) Peak-to-peak THz amplitude as a function of thickness (symbol: measured data; solid-line: fitting using Eq. (1)).

FIG. 4. Pumping fluence dependence of THz peak amplitude from the FeMnPt single-layer FM emitter: Peak amplitude as a function of fluence for emitters with different FeMnPt thickness. Inset: coefficient ($\gamma$) of laser heating effect on saturation magnetization.

FIG. 5. (a) THz emission from FeMnPt with uniform Pt composition. (b) THz emission from FeMnPt with positive Pt composition gradient. (c) THz emission from FeMnPt with negative Pt composition gradient. In all the figures, the upper pair of waveforms is from quartz side pumping and lower pair is from MgO side pumping. Solid and dashed-lines correspond to positive and negative fields, respectively. Darker color indicates the region with higher Pt composition.

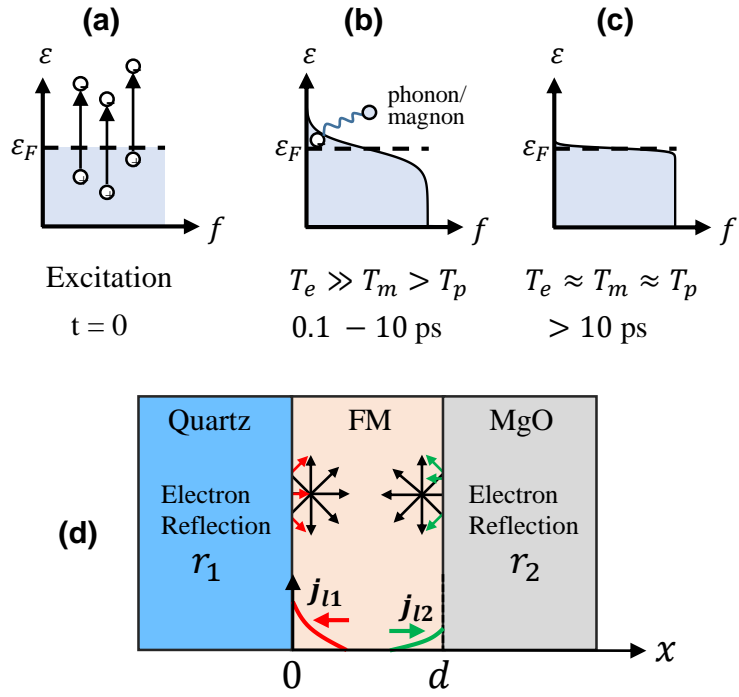

FIG. 1.

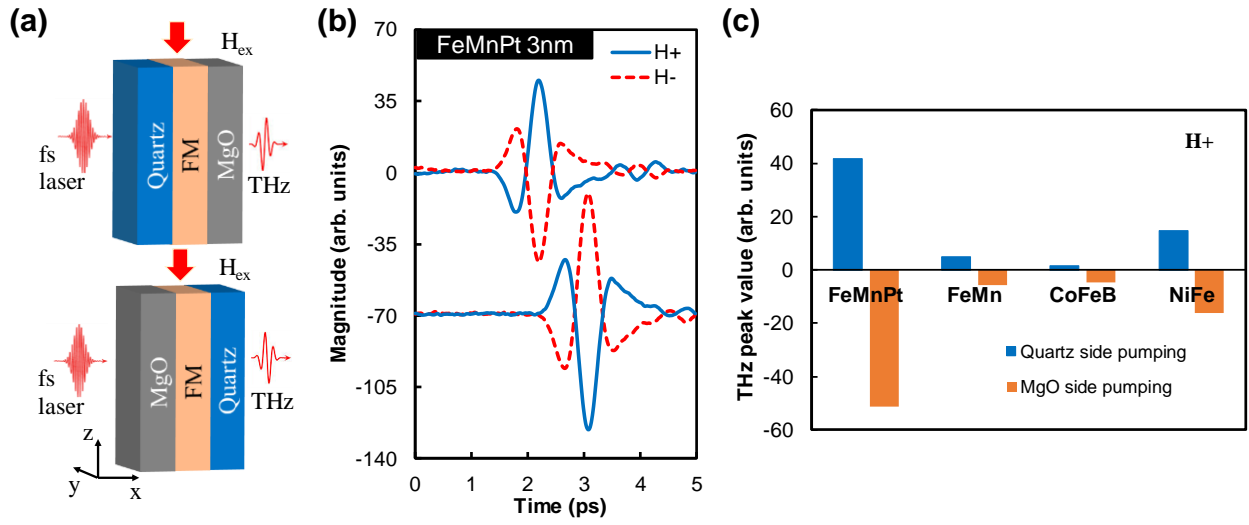

FIG. 2.

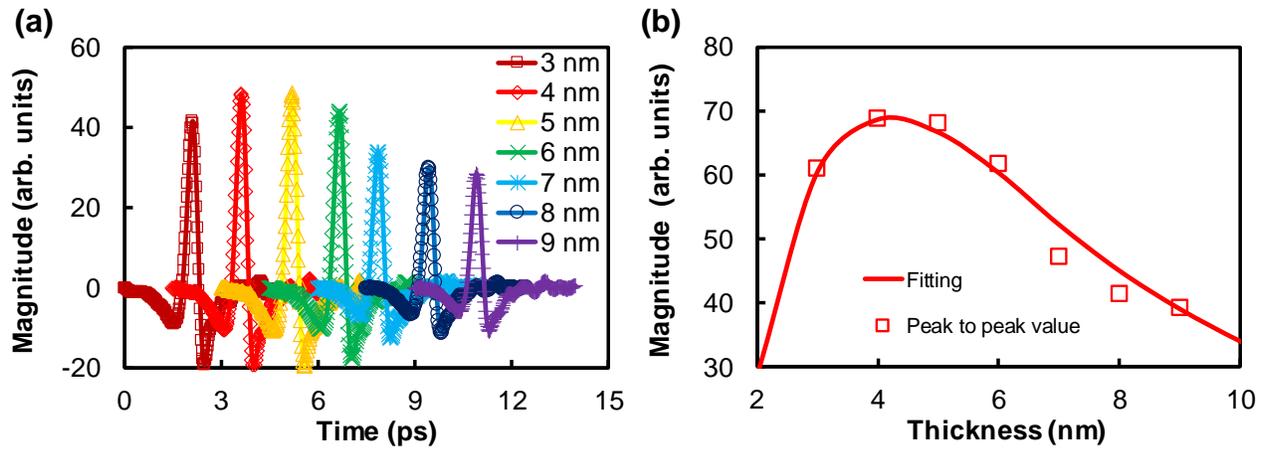

FIG. 3.

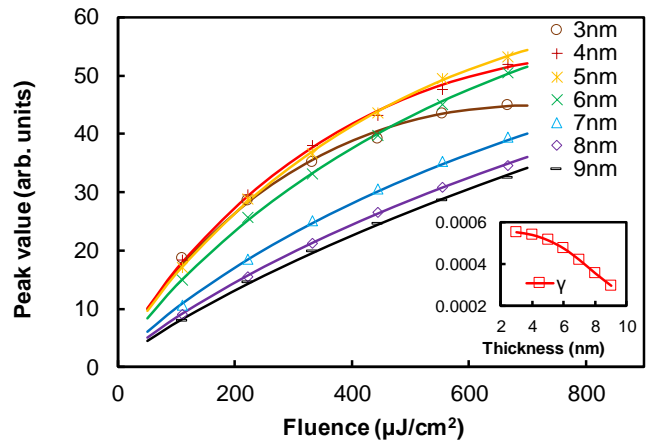

FIG. 4.

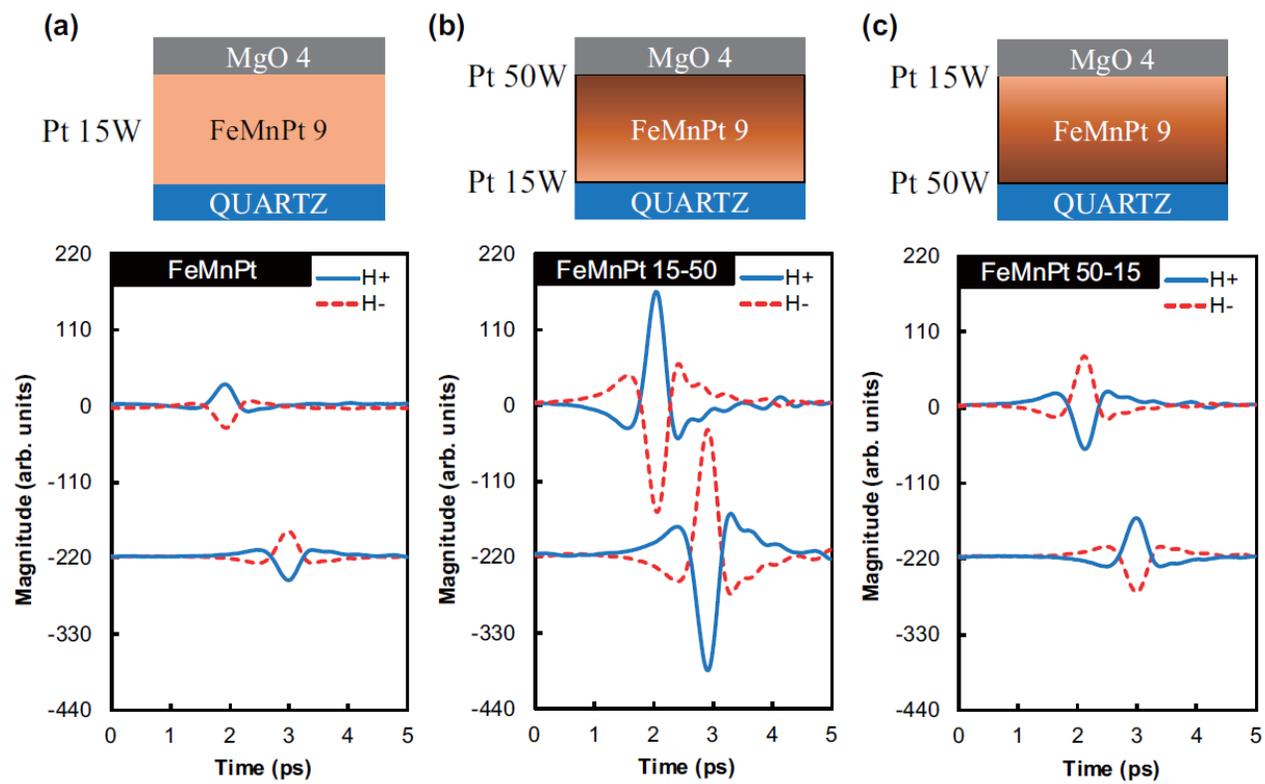

FIG. 5.